\documentclass[structabstract]{aa}  
\usepackage{graphicx}
\usepackage{txfonts}
\usepackage{multirow}
\usepackage[]{natbib}
\bibpunct{(}{)}{;}{a}{}{,}
\begin{document}
   \title{\object{GRB\,090313}: X-shooter's first shot at a GRB\thanks{Based on observations made with ESO VLT at Paranal Observatory during the commissioning of X-shooter.}}

   \author{
   	A.~de~Ugarte Postigo \inst{1,2}
          \and
          P.~Goldoni \inst{3,4}
          \and
          C.C.~Th\"one \inst{1}
          \and
          S.D.~Vergani \inst{3}
          \and
          V.~D'Elia \inst{5,6}
          \and
          S.~Piranomonte \inst{5}
          \and
          D.~Malesani \inst{7}
          \and
          S.~Covino \inst{1}
          \and
          H.~Flores \inst{8}
          \and
          J.P.U.~Fynbo \inst{7}
          \and
          J.~Hjorth \inst{7}
          \and
          R.A.M.J.~Wijers \inst{9}
          \and
          S.~D'Odorico \inst{10}
          \and
          F.~Hammer \inst{8}
          \and
          L.~Kaper \inst{9}
          \and
          P.~Kj\ae rgaard \inst{11}
          \and
          S.~Randich \inst{12} 
          \and
          M.I.~Andersen \inst{7}
          \and
          L.A.~Antonelli \inst{5}
          \and
          L.~Christensen \inst{10}
          \and
          P.~D'Avanzo \inst{1}
          \and
          F.~Fiore \inst{5}
          \and
          P.J.~Groot \inst{13}
          \and
          E.~Maiorano \inst{14}
          \and
	 E.~Palazzi \inst{14}
          \and
          E.~Pian \inst{15,16, 10} 
          \and
          G.~Tagliaferri \inst{1}
          \and
          M.E.~van~den~Ancker \inst{10}
          \and
          J.~Vernet \inst{10}
          }
   \institute{
                  INAF - Osservatorio Astronomico di Brera, via E. Bianchi 46, 23807, Merate, Lc, Italy.
       \and   European Southern Observatory, Casilla 19001, Santiago 19, Chile.
       \and   APC - UMR 7164, 10 rue Alice Domon et Leonie Duquet 75205 Paris Cedex 13 France.
       \and   Service D'Astrophysique, DSM/IRFU/SAp, CEA-Saclay, 91191, Gif-sur-Yvette France.
       \and   INAF - Osservatorio Astronomico di Roma, Via Frascati, 33, I-00040, Monteporzio Catone (Rome), Italy.
       \and   ASI - Science Data Center, Via Galileo Galilei, 00044, Frascati (Rome) Italy.
       \and   Dark Cosmology Centre, Niels Bohr Institute, University of Copenhagen, Juliane Maries Vej 30, 2100 Copenhagen \O, Denmark.
       \and   GEPI, Observatoire de Paris, CNRS, Universit\'e Paris Diderot; 5 Place Jules Janssen, Meudon, France.
       \and   Astronomical Institute, University of Amsterdam, Kruislaan 403, 1098 SJ Amsterdam, The Netherlands.
       \and   European Southern Observatory, K. Schwarzschild Str. 2, 85748 Garching, Germany.
       \and   Niels Bohr Institute, University of Copenhagen, Juliane Maries Vej 30, 2100 Copenhagen \O, Denmark.
       \and   INAF - Osservatorio Astrofisico di Arcetri, Largo E. Fermi 5, 50125 Firenze, Italy.
       \and   Department of Astrophysics, IMAPP, Radboud University Nijmegen, P.O. Box 9010, 6500 GL Nijmegen, The Netherlands.
       \and   INAF - IASF di Bologna, via Gobetti 101, I-40129 Bologna, Italy.
       \and   INAF - Osservatorio Astronomico di Trieste, via Tiepolo, 11, 34131 Trieste, Italy.
       \and   Scuola Normale Superiore di Pisa, Piazza dei Cavalieri 7, I-56126 Pisa, Italy.
             }

   \date{Received; accepted}

  \abstract
   {X-shooter is the first second-generation instrument to become operative at the ESO Very Large Telescope (VLT). It is a broad-band medium-resolution spectrograph designed with gamma-ray burst (GRB) afterglow spectroscopy as one of its main science drivers.}
   {During the first commissioning night on sky with the instrument fully assembled, X-shooter observed the afterglow of GRB\,090313 as a demonstration of the instrument's capabilities.}
   {GRB\,090313 was observed almost two days after the burst onset, when the object had already faded to \textit{R}$\sim$21.6. Furthermore, the 90\% illuminated Moon was just 30 degrees away from the field. In spite of the adverse conditions, we obtained a spectrum that, for the first time in GRB research, covers simultaneously the range from 5\,700 to 23\,000$~\AA$.}
   {The spectrum shows multiple absorption features at a redshift of 3.3736, which we identify as the redshift of the GRB. These features are composed of 3 components with different ionisation levels and velocities. Some of the features have never been observed before in a GRB at such a high redshift. Furthermore, we detect two intervening systems at redshifts of 1.8005  and 1.9597.}
   {These results demonstrate the potential of X-shooter in the GRB field, as it was capable of observing a GRB down to a magnitude limit that would include 72\% of long GRB afterglows 2 hours after the burst onset. Coupled with the rapid response mode available at VLT, allowing reaction times of just a few minutes, X-shooter constitutes an important leap forward on medium resolution spectroscopic studies of GRBs, their host galaxies
and intervening systems, probing the early history of the Universe.}


   \maketitle


\section{Introduction}

During the first hours after the onset of long gamma-ray bursts \citep[GRBs, for a short-long classification see][]{kou93}, their optical/near infrared (nIR) counterparts shine as the brightest beacons in the Universe \citep{kan07, rac08, blo09}. Being generally produced by the collapse of a massive star \citep[][and references therein]{woo06}, they give us the opportunity to study the environment of massive star forming regions at cosmological distances \citep[the average redshift of long GRBs observed by \textit{Swift} is $\sim$2.2;][]{fyn09} through the use of medium to high resolution spectroscopy. Time variability of absorption features corresponding to fine-structure and metastable level transitions has been sometimes identified \citep{des06,vre07,del09}, which makes it possible to constrain the distance from the GRB to the absorbing gas, at typical scales of kpc. Only few such cases have been spotted up to now, due the required spectral resolution and sensitivity. Furthermore, being bright sources that disappear over time, they allow us to study intervening systems through their absorption signatures in the afterglow spectrum, and to search for the associated galaxies when the afterglow has faded away \citep{jak04, che09}.

The family of short GRBs, although still poorly studied, presents on average fainter counterparts and has been suggested to be produced by the coalescence of compact objects \citep[][and references therein]{nak07}, although some of them could come from other progenitors, such as extragalactic magnetars \citep{hur05}. Up to now, there have been few attempts to acquire a spectrum of a short GRB, most of them having been unsuccessful due to late observations or faint afterglows. There is still no medium or high resolution spectrum of a short GRB.

Within this context, X-shooter \citep{dod06} is presented to the GRB community. It is the first of the second-generation instruments at ESO's Very Large Telescope (VLT), at Paranal Observatory (Chile). It is a single target spectrograph capable of obtaining a medium resolution spectrum ($\Re$=4\,000 - 14\,000, depending mainly on the slit width and wavelength) covering the complete range from 3\,000 to 24\,800 $\AA$ in a single exposure. It has been designed to maximise efficiency by splitting the light with dichroics into three arms: ultraviolet/blue (UVB), visible (VIS) and near-infrared (NIR). Each arm has an echelle spectrograph with optimised optics, coatings, dispersive elements and detectors. Since March 2009, it is installed in the Cassegrain focus of \textit{Kueyen}, the second 8.2m Unit Telescope of the VLT, where it began routine scientific observations in October 2009. 

One of the key science drivers of X-shooter is the study of optical and nIR afterglows of GRBs as cosmological probes of the circumstellar, interstellar and intergalactic medium back to the epoch of first star formation in the Universe. During its first commissioning night on sky with its full setup, X-shooter was used to observe the afterglow of GRB\,090313, which had exploded two days before, as a test target. In this article we present the results of this observation of a GRB with X-shooter, giving a hint of the full potential that the instrument will have during regular observations.

The paper is organised as follows: Sect. 2 describes our observations in the context of the GRB evolution. In Sect. 3 we present our results, focusing on the spectral features local to the GRB and on the intervening systems. Finally, in Sect. 4 we discuss the results and present our conclusions.

\section{Observations}

At 09:06:27 UT on 13 March 2009, \textit{Swift}'s Burst Alert Telescope (BAT) detected GRB\,090313 \citep{mao09}, a long-duration burst with $T_{90} =78\pm19$ s \citep{sak09}. In response to the trigger, the 0.76m Katzman Automatic Imaging Telescope (KAIT) slewed to the position shortly after and detected a bright ($R$$\sim$16 magnitude) optical afterglow \citep{cho09a} at equatorial coordinates (J2000.0): R.A. = $13^\mathrm{h}13^\mathrm{m}36\fs21$, Dec. = $+08^{\circ}05^{\prime}49\farcs2$ \citep{upd09}. 

The afterglow evolved through a plateau phase \citep{per09a,deu09a} during which it maintained a magnitude of $I$$\sim$17.7 up to almost a day after the burst, when the light curve steepened \citep{deu09b} and decayed following a power law $F\propto t^{-1.77}$  \citep{per09b}. \citet{per09c} used multicolour photometry to study the spectral energy distribution of the afterglow, concluding that its intrinsic extinction followed a Small Magellanic Cloud profile with a $V$-band extinction $A_V$ $\lesssim 0.4$. This value was later refined by \citet{kan09} to $A_V$ = 0.34 $\pm$ 0.15. They note that in the time interval between 0.02 and 0.5 days the afterglow of GRB\,090313 was the optically brightest ever detected. A detailed multiwavelength study of the afterglow will be presented by Melandri et al. (in preparation).

\begin{figure}[h!]
   \centering
   \includegraphics[width=8cm]{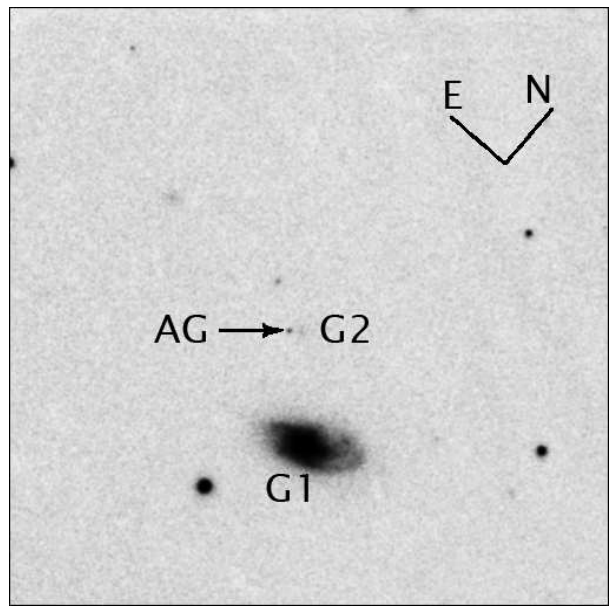}
      \caption{\textit{I}-band acquisition image. The GRB afterglow is indicated with an arrow. We have also marked the two galaxies identified by \citet{ber09}, mentioned in the discussion, as G1 and G2 (G2 is faintly visible $\sim 2$\arcsec{} NW of the afterglow). The figure shows the complete field of view of the acquisition and guiding camera, $90^{\prime\prime}\times90^{\prime\prime}$. North and East are indicated in the figure.
              }
         \label{FigAcq}
   \end{figure}

Shortly after X-shooter came online for commissioning, it was aimed at the afterglow of GRB\,090313. The observations consisted of a 900 s exposure and four additional exposures of 1500 s each, with a slit width of 0\farcs9 in the VIS and NIR arms and 1\farcs0 in UVB. The UVB and VIS detectors were used with 1$\times$2 binning (binned in the spectral direction but not in the spatial one) and a slow readout of 100 kHz to minimise the noise. The NIR detector was used in the default unbinned sample-up-the-ramp (non-destructive) mode. The resolution of the final spectrum varies with wavelength from 35 to 60 km s$^{-1}$. All three spectral ranges were exposed during the first 900 s. Inspection of this first set of frames showed a negligible signal in the UVB arm due to strong contamination by the Moon, which was just 30 degrees away from the field and almost full (90\%). Hence, for the subsequent exposures, only VIS and NIR frames were obtained. The mid-exposure time of the combined spectrum is 15.26 March 2009 UT, 45.1 hours after the onset of the burst, when the afterglow had already faded to \textit{R}$\sim$21.6 \citep{per09b, cob09}.  On a 2$\times$20 s combination of acquisition frames obtained at a mean time of 43.82 hours after the burst we measure \textit{I}=20.63$\pm$0.06 (see Fig.~\ref{FigAcq}). This magnitude is based on Cousin magnitudes derived from the SDSS catalogue\footnote{http://www.sdss.org/DR7/algorithms/sdssUBVRITransform.html}. The signal to noise ratio (S/N) per spectral bin of the spectrum varies with wavelength, reaching a maximum of $\sim$7 around 8\,000 $\AA$.

\begin{figure*}[]
   \centering
   \includegraphics[width=17cm]{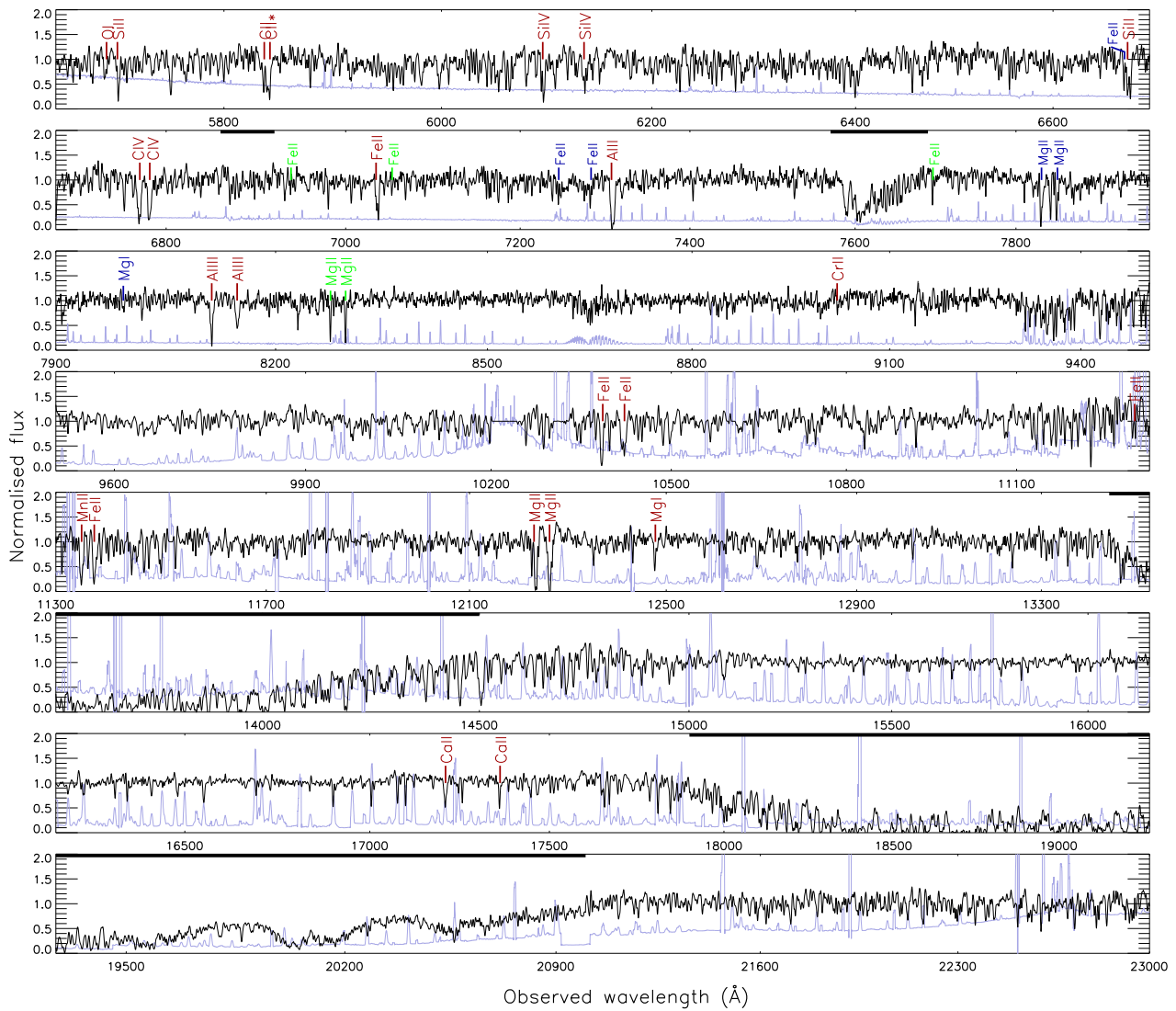}
      \caption{Normalised spectrum of the afterglow of GRB\,090313, smoothed with a Gaussian filter (using a smoothing kernel with full width half maximum varying from 1 to 4 \AA) for displaying purposes. The absorption features produced at the redshift of the GRB afterglow ($z=3.3736$) are indicated in larger font size (red). The intervening systems at $z=1.8005$ and $z=1.9597$ are indicated in smaller case and using dark (blue) and light (green) traces respectively. The thick horizontal lines at the top of each panel indicate regions affected by strong telluric bands, which have been excluded from the analysis. The error spectrum is marked as a pale (blue) line. See the electronic edition for a color version of this figure.
              }
         \label{FigSpec}
   \end{figure*}

The data were reduced using a standard procedure with a preliminary version of X-shooter's pipeline \citep{gol06,gol08}. Due to variable conditions during the night of the observation, the spectra were not flux calibrated and a normalised spectrum is used in this work. No telluric correction was applied, so that prominent atmospheric bands (especially in the NIR) can be still seen in Fig.~\ref{FigSpec}, where we show the reduced normalised spectrum, which is a combination of all the individual spectra, covering the useful observing range (where the afterglow was detectable above the background level). The wavelengths reported throughout the paper are in vacuum. The uncertainty in the wavelength calibration is of the order of 30 km s$^{-1}$ when using the current pipeline, which is the dominant uncertainty when calculating the redshift. For the analysis of the spectral features we have used FITLYMAN \citep{fon95}. 

We caution that, being the first night of commissioning on sky, the instrument was not yet optimised. Thus, the results presented here must not be considered as representing the optimal performance of the instrument but as a preliminary view of the potential that will be delivered by the instrument in regular operation. The results presented here  supersede those reported by \citet{deu09c}.

\section{Results}

Over the complete range of the spectrum we identify a main absorption system that is composed of several velocity components. In order to determine the central redshift of the absorption system, the spectra were rebinned such that the resulting absorption line centre represents the wavelength where the strongest absorption takes place. Excluding from our calculations the blended absorption lines \ion{C}{II}$\lambda\lambda$1334 and \ion{C}{II*}$\lambda$1335.6, 1335.7 (see Sect.~3.1), we derive a redshift of 3.3736$\pm$0.0004 which we identify as the redshift of the GRB. This value is consistent with the one previously determined by \citet{cho09b} using GMOS/Gemini-South and \citet{tho09} using FORS2/VLT. 

Furthermore, we identify two intervening systems in the spectrum, one of which can be again divided into several velocity components. We determine the redshift in the same way as for the host galaxy system and derive redshifts of 1.9597$\pm$0.0003 for the higher redshift system and  1.8005$\pm$0.0003 for the multi-component lower redshift system. Fig.~\ref{FigSpec} shows the complete normalized X-shooter spectrum, where the absorption features of both the GRB and the intervening absorption systems are marked. The rest frame equivalent widths (EW$_{\mathrm{rest}}$) of all identified features are listed in Table~\ref{table:ewlines}, together with their observed and rest frame wavelengths and redshift.

\begin{table}[h]
\caption{List of features identified in the spectra (detections $\gtrsim$3-$\sigma$) along with their rest-frame equivalent widths. All wavelengths are in vacuum. For homogeneity, the values for all the velocity components are combined together, as only some of them could be measured independently. We have also divided the table into three parts with the host redshift lines, intervening system at \textit{z}=1.96 and intervening system at \textit{z}=1.80.}

\label{table:ewlines}      
\centering                          
\begin{tabular}{c c c c}        
\hline\hline                 
\textbf{$\lambda_{\mathbf{obs}}$}  & \textbf{Feature}                    & \textbf{$z$} & \textbf{EW$_{\mathbf{rest}}$}            \\    
{\scriptsize \textbf{(\AA)}} & {\scriptsize \textbf{(\AA)} }  &                          & {\scriptsize \textbf{(\AA)} }   \\  \hline\hline  
\multicolumn{4}{c}{Host galaxy at $z=3.37$}\\
\hline 
5695.71	& \ion{O}{I}$\lambda$1302.17   & 3.37402	&  0.36$\pm$0.11\\		
5705.96	& \ion{Si}{II}$\lambda$1304.37 &	3.37450	&  0.58$\pm$0.12\\	
                                               & \ion{C}{II}$\lambda$1334.53  &  & \\
5839.99	& \ion{C}{II*}$\lambda$1335.66  & 3.37413 &   0.88$\pm$0.12 \\
                                               & \ion{C}{II*}$\lambda$1335.71  &  & \\
6096.13  & \ion{Si}{IV}$\lambda$1393.75  & 3.37389    & 1.14$\pm$0.13  \\
6135.08  & \ion{Si}{IV}$\lambda$1402.77  & 3.37354    & 1.03$\pm$0.10  \\
6678.24  & \ion{Si}{II}$\lambda$1526.71    &  3.37428   &  0.95$\pm$0.08 \\
6770.31  & \ion{C}{IV}$\lambda$1548.19   &  3.37303  & 0.91$\pm$0.07  \\
6781.67  & \ion{C}{IV}$\lambda$1550.77   &  3.37310   & 0.89$\pm$0.06  \\
7035.95  & \ion{Fe}{II}$\lambda$1608.45   &  3.37437 & 0.62$\pm$0.06  \\
7308.47  & \ion{Al}{II}$\lambda$1670.79    & 3.37427  & 1.29$\pm$0.06  \\
8111.69  & \ion{Al}{III}$\lambda$1854.72   &  3.37355  & 0.71$\pm$0.04 \\
8146.86  & \ion{Al}{III}$\lambda$1862.79   &  3.37347  & 0.56$\pm$0.05  \\
9019.61  &\ion{Cr}{II}$\lambda$2062.23    & 3.37371   & 0.14$\pm$0.02\\
10385.9  & \ion{Fe}{II}$\lambda$2374.46   & 3.37401 &  1.12$\pm$0.10   \\
10420.9  & \ion{Fe}{II}$\lambda$2382.77   & 3.37348  & 1.11$\pm$0.17  \\
11312.1  &\ion{Fe}{II}$\lambda$2586.65    & 3.37326  & 1.21$\pm$0.49\\
11347.5  & \ion{Mn}{II}$\lambda$2594.74  & 3.37327  & 1.04$\pm$0.14  \\
11372.2  & \ion{Fe}{II}$\lambda$2600.17   & 3.37363  & 1.29$\pm$0.11 \\
12233.7  & \ion{Mg}{II}$\lambda$2796.35  & 3.37486  & 1.94$\pm$0.15  \\
12267.9  & \ion{Mg}{II}$\lambda$2803.53  & 3.37588  & 1.90$\pm$0.13  \\
12476.5  & \ion{Mg}{I}$\lambda$2852.96   & 3.37317 & 0.51$\pm$0.05  \\
17207.2	& \ion{Ca}{II}$\lambda$3934.78 & 3.37311  & 1.21$\pm$0.06 \\
17359.6  & \ion{Ca}{II}$\lambda$3969.59 & 3.37314  & 0.45$\pm$0.04 \\
\hline
\multicolumn{4}{c}{Intervening system at $z= 1.96$}\\
\hline 
6937.93  & \ion{Fe}{II}$\lambda$2344.21  & 1.95960  & 0.15$\pm$0.04  \\
7052.26  & \ion{Fe}{II}$\lambda$2382.77  & 1.95969  & 0.18$\pm$0.04  \\
7695.64  & \ion{Fe}{II}$\lambda$2600.17  & 1.95966  & 0.31$\pm$0.04  \\
8276.48  & \ion{Mg}{II}$\lambda$2796.35 & 1.95974  & 0.50$\pm$0.03  \\
8297.48  & \ion{Mg}{II}$\lambda$2803.53 & 1.95965 & 0.59$\pm$0.05  \\ 
\hline
\multicolumn{4}{c}{Intervening system at $z= 1.80$}\\
\hline 
6673.45  & \ion{Fe}{II}$\lambda$2382.77 & 1.80072    & 0.34$\pm$0.05  \\
7240.68  & \ion{Fe}{II}$\lambda$2586.65 & 1.79925    & 0.48$\pm$0.06  \\
7280.21  & \ion{Fe}{II}$\lambda$2600.17 & 1.79989 & 0.67$\pm$0.07  \\
7830.87  & \ion{Mg}{II}$\lambda$2796.35 & 1.80039  & 1.35$\pm$0.10  \\
7849.07  & \ion{Mg}{II}$\lambda$2803.53 & 1.79970 & 1.31$\pm$0.08  \\
7987.58  & \ion{Mg}{I}$\lambda$2852.96 & 1.79974 & 0.48$\pm$0.08 \\

\hline\hline
\end{tabular}
\end{table}

\subsection{The host galaxy system}

\begin{figure}[h!]
   \centering
  \includegraphics[angle=0,width=8cm]{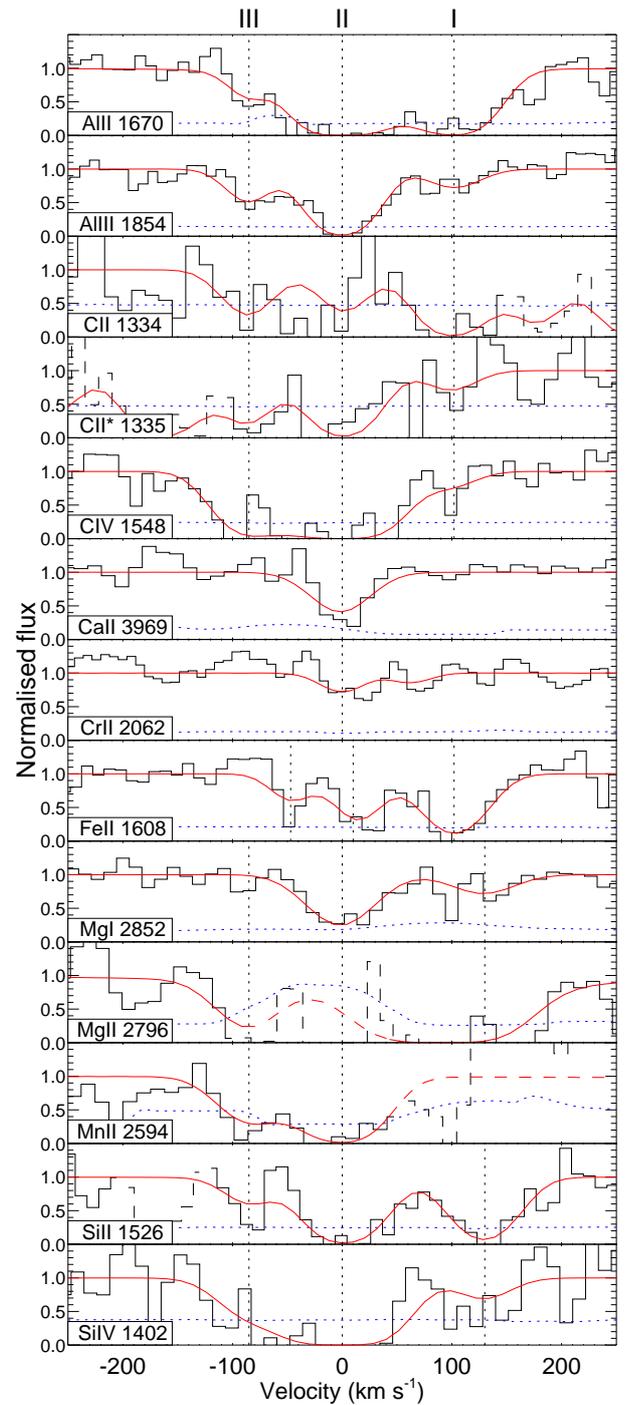}
      \caption{Fit of the detected of the detected transitions at the GRB redshift. For the elements with several lines detected, we show the transitions that are least saturated or have the highest S/N. The regions of the spectrum that are excluded from the fit due to contamination by sky lines or other unrelated absorption features are drawn with a dashed line. We indicate the error spectrum with a dotted (blue) line. Vertical dotted (black) lines mark the different velocity components. See the electronic edition for a color version of this figure.
              }
         \label{FigGRB}
   \end{figure}

Most of the absorption features detected at the redshift of the GRB can be separated into three different components that we name I, II and III, with I having the highest redshift. We fit line profiles to the different components in order to determine the column density of each species. Due to the limited resolution and S/N, we fix the thermal $b$-parameter to 5 km~s$^{-1}$ and the turbulent $b$-parameters to 20 km~s$^{-1}$ (values that give the best global result for the fits), leaving only the column density as a free parameter. Transitions of the same element and ionization level were fitted with a common column density.  Some transitions are not well fitted with only three components, which might indicate that there are other, smaller components not resolved in our spectrum. Furthermore, some of the lines are clearly saturated, so that only lower limits of the column densities can be derived for them. The results of the fits are displayed in Table~\ref{table:lines}. The fits to the highest S/N or least saturated transitions for each ion are shown in Fig.~\ref{FigGRB}.

The exact number of components for \ion{Mg}{II} is not known due to the contamination with skylines and we therefore fitted just the outer components I and III. Moreover, due  to saturation, we could derive only lower limits for the column densities of these components. For \ion{Mn}{II} we could not fit component I due to contamination by skylines. The absorption bluewards of component III in \ion{Si}{II}$\lambda$1526 is due to \ion{Fe}{II}$\lambda$2382 from the intervening system at redshift 1.80. We did not fit the OI $\lambda$ 1302 transition due to the low S/N in the blue end of the spectrum. For weaker lines where we do not detect all three components, we give 3-$\sigma$ upper limits for those components unless those regions are affected by skylines or other absorption transitions.

The central component II is the one with the highest column density and except for \ion{Fe}{II} it could be fixed to 0\,km s$^{-1}$. The blueshifted component III is around --85\,km s$^{-1}$ for all elements where this component is present except for \ion{Fe}{II} where it lies at --47 \,km s$^{-1}$.  For component I, the situation is slightly more complex. \ion{Al}{II}, \ion{Al}{III} and \ion{Fe}{II} have component I at +102 \,km s$^{-1}$ whereas it is at +130\,km s$^{-1}$ for the Si lines as well as for \ion{Mg}{II}. Components I and III are absent for the weak lines \ion{Ca}{II}, \ion{Cr}{II} and \ion{Zn}{II}. For \ion{Mg}{I} we only detect components I and II whereas for \ion{C}{IV} we only detect components II and III. \ion{Fe}{II} has 3 components but at slightly different velocities than the other lines. We cannot confirm this behaviour securely by fitting other \ion{Fe}{II} lines as the $\lambda$ 1608 \AA{} line is the only one not severely affected by skylines.  This behaviour of the different components might indicate that the absorbing material is not at the same place along the line-of-sight for different elements, which is expected if the absorption lines trace gas at different locations inside the host galaxy. Generally, we find that in component I the low ionization species are predominant over the higher ionization species, while for component III the opposite is true. This is also confirmed by the ratio of column densities of \ion{Al}{III} and \ion{Al}{II} for which we get values of 0.03$\pm$0.03, 0.14$\pm$0.13 and 4.2$\pm$0.9  for components I, II and III respectively. A similar behaviour is observed for the ratio of column densities of \ion{Si}{IV} and \ion{Si}{II} with values of 0.02$\pm$0.01, 0.91$\pm$0.26 and 1.18$\pm$0.14 for components I to III.

We detect the fine structure transitions of \ion{C}{II}$\lambda$1334, \ion{C}{II*}$\lambda$1335.6, 1335.7. The latter two have only a 0.1 $\AA$ difference in central wavelength,  and therefore cannot be separated with the spectral resolution provided by X-shooter. However, the 1335.7 $\AA$ transition is 10 times stronger than the 1335.6 $\AA$ one and we therefore fit the absorption systems assuming that the stronger transition is responsible for most of the absorption. The resonant and fine structure transitions are separated well enough to allow a fit of all 3 components for both lines, only component III of the fine structure transition is rather close to component I of the resonant line, making the system blended. On the other hand we cannot identify any \ion{Si}{II*} lines at the redshift of the GRB, which were reported for this GRB by \citet{cho09b}. However, our spectrum only covers the 1533.4 $\AA$ and the 1309.2 $\AA$ features and not that at 1264.7 $\AA$ which is significantly stronger. We determine a 3-$\sigma$ upper limit for \ion{Si}{II*}$\lambda$1533.4 of EW$_{\mathrm{rest}}<$0.21 $\AA$ and a limit for \ion{Si}{II*}$\lambda$1309.2 of EW$_{\mathrm{rest}}<$0.58 $\AA$, the first being more constraining due to both the higher oscillator strength and signal to noise ratio.

\begin{table*}
\caption{List of features identified in the spectrum (detections $\gtrsim$3-$\sigma$). Most lines show three velocity components. Transitions in brackets are contaminated by skylines or blended and have not been used to fit the column densities. Upper limits are 3-$\sigma$ where the wavelengths were fixed to the corresponding components observed in other transitions. For saturated features we give lower limits for the column density.}             
\label{table:lines}      
\centering                          
\begin{tabular}{c c c c c c c c}        
\hline\hline                 
 &  & \multicolumn{2}{c}{\textbf{Component I}}& \multicolumn{2}{c}{\textbf{Component II}} & \multicolumn{2}{c}{\textbf{Component III}} \\    
\hline
    \textbf{Ion}   &     \textbf{Transitions}           & \textbf{Velocity} &\textbf{log N }  & \textbf{Velocity }&\textbf{log N }  &\textbf{Velocity }&\textbf{log N }  \\ 
       &\scriptsize \textbf{ (\AA)}		&\scriptsize \textbf{($\mathbf{km~s^{-1}}$)}&\scriptsize \textbf{($\mathbf{cm^{-2}}$)}&\scriptsize \textbf{($\mathbf{km~s^{-1}}$)}&\scriptsize \textbf{($\mathbf{cm^{-2}}$)}&\scriptsize \textbf{($\mathbf{km~s^{-1}}$)}&\scriptsize \textbf{($\mathbf{cm^{-2}}$)}\\
\hline\hline  
 \ion{Si}{IV} & 1393, 1402 & +130 &13.3$\pm$0.3    & 0 & $>$15.3& --85 & 13.8$\pm$0.3\\ %
 \ion{Si}{II}  & 1526           & +130 &15.0$\pm$0.5 & 0 & $>$15.4 &--85	&13.7$\pm$0.2 \\
\ion{C}{IV}  & 1548, 1550 &+102 &13.2$\pm$0.3  &0& $>$15.2&--85&$>$14.5\\ %
\ion{C}{II}  & 1334  & +102 &  14.1$\pm$0.4  &0&14.2$\pm$0.4 &--85&15.3$\pm$0.6\\ %
\ion{C}{II*}  & 1335  &+102 &  15.3$\pm$0.8  &0&14.4$\pm$0.4 &--85&13.7$\pm$0.5\\ %
\multirow{2}{*}{\ion{Fe}{II}}  & 1608, (2374) & \multirow{2}{*}{+102} &\multirow{2}{*}{14.9$\pm$0.2} & \multirow{2}{*}{+10} & \multirow{2}{*}{14.5$\pm$0.1} & \multirow{2}{*}{--47} & \multirow{2}{*}{13.9$\pm$0.2} \\ %
                     & (2382, 2585, 2600) & & &&&& \\
\ion{Al}{III}  & 1854,1862  & +102 &12.8$\pm$0.1 & 0& 14.4$\pm$0.2 & --85 & 13.1$\pm$0.1\\ %
\ion{Al}{II}   & 1670            & +102 &$>$14.6& 0 & $>$15.4 & --85 & 12.5$\pm$0.2  \\ %
\ion{Cr}{II} & 2062  &--- &  ---  & 0 & 13.6$\pm$0.1     & (--85)  & $<$13.6            \\ %
\ion{Mg}{II} & 2796, (2803)  &+130& $>$14.8         &--- &---                   &--85 & $>$13.6   \\
\ion{Mg}{I}  & 2853            & +130&12.2$\pm$0.3        & 0 &13.1$\pm$0.1 &(--85)  &$<$12.6        \\ %
\ion{Mn}{II}  & 2594           & ---&---        & 0 &$>$14.3&--85  &$>$13.6        \\ %
\ion{Ca}{II} & (3933), 3969 & (+102) & $<$12.6         & 0 & 13.3$\pm$0.4             & (--85) & $<$13.0  \\ %
\hline\hline
\end{tabular}
\end{table*}   
   
\begin{table*}
\caption{List of column densities measured for the transitions identified in the $z$=1.96 intervening system.}             
\label{table:int1}      
\centering                          
\begin{tabular}{c c c c}   
\hline\hline  
\multicolumn{2}{c}{\textbf{}}  & \multicolumn{2}{c}{\textbf{Component I}}\\    
\hline
    \textbf{Ion}   &     \textbf{Transitions}           & \textbf{Velocity} &\textbf{log N }     \\ 
       &\scriptsize \textbf{ (\AA)}		&\scriptsize \textbf{($\mathbf{km~s^{-1}}$)}&\scriptsize \textbf{($\mathbf{cm^{-2}}$)}\\
\hline\hline  
\ion{Fe}{II}  & 2344, 2382, 2600 & 0 & 14.9$\pm$0.5   \\ %
\ion{Mg}{II} & 2796,2803 & 0 &  $>15.5$   \\ %
\hline\hline
\end{tabular}
\end{table*}

\begin{table*}
\caption{List of column densities measured for the transitions identified in the $z$=1.80 intervening system. This system presents 3 different velocity components.}             
\label{table:int2}      
\centering                          
\begin{tabular}{c c c c c c c c}   
\hline\hline                 
\multicolumn{2}{c}{\textbf{}}  & \multicolumn{2}{c}{\textbf{Component I}}& \multicolumn{2}{c}{\textbf{Component II}} & \multicolumn{2}{c}{\textbf{Component III}} \\    
\hline
    \textbf{Ion}   &     \textbf{Transitions}           & \textbf{Velocity} &\textbf{log N }  & \textbf{Velocity }&\textbf{log N }  &\textbf{Velocity }&\textbf{log N }  \\ 
       &\scriptsize \textbf{ (\AA)}		&\scriptsize \textbf{($\mathbf{km~s^{-1}}$)}&\scriptsize \textbf{($\mathbf{cm^{-2}}$)}&\scriptsize \textbf{($\mathbf{km~s^{-1}}$)}&\scriptsize \textbf{($\mathbf{cm^{-2}}$)}&\scriptsize \textbf{($\mathbf{km~s^{-1}}$)}&\scriptsize \textbf{($\mathbf{cm^{-2}}$)}\\
\hline\hline 
\ion{Fe}{II}  & 2382, 2586, 2600 & +85 & 13.3$\pm$0.1 & 0 & 15.0$\pm$0.2 & --286 & 14.0$\pm$0.1 \\ %
\ion{Mg}{II} & 2796, 2803 & +85 &13.4$\pm$0.1 & 0 &$>15.3$& --286 &$>14.7$ \\ %
\ion{Mg}{I}  & 2853           & +85 & 12.3$\pm$0.2 & 0 & $12.0\pm0.2$ & --286 & 12.5$\pm$0.2\\ %
\hline\hline
\end{tabular}
\end{table*}

Combining the column density of all three components we find a value of [Si/Fe] $ > 0.65$ for the GRB host galaxy \citep[reference Solar composition obtained from][]{asp05}. This large value of [Si/Fe] is in agreement with the work of \citet{pro07} for GRBs, being higher than those found for intervening absorbers in quasar spectra.

\subsection{Intervening systems}

\begin{figure}[h]
   \centering
   \includegraphics[width=8cm]{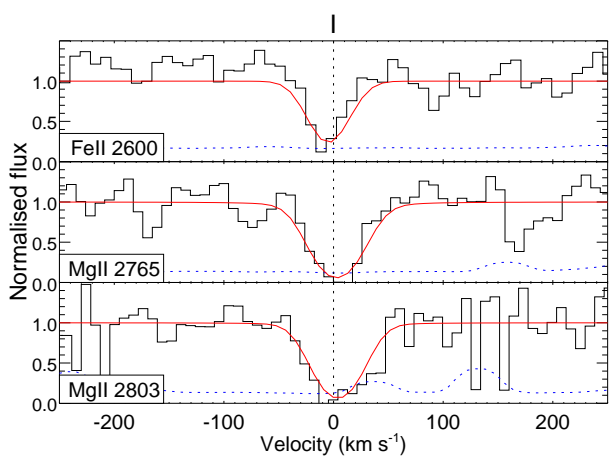}
      \caption{Fit of \ion{Mg}{II} and \ion{Fe}{II} lines of the intervening system at $z$=1.9597.
              }
         \label{Fig1.9}
   \end{figure}

An intervening system is identified at a redshift of 1.9597 through the detection of \ion{Fe}{II} and \ion{Mg}{II} in absorption (see Fig.~\ref{Fig1.9} and Table~\ref{table:int1}).

\begin{figure}[h]
   \centering
   \includegraphics[width=8cm]{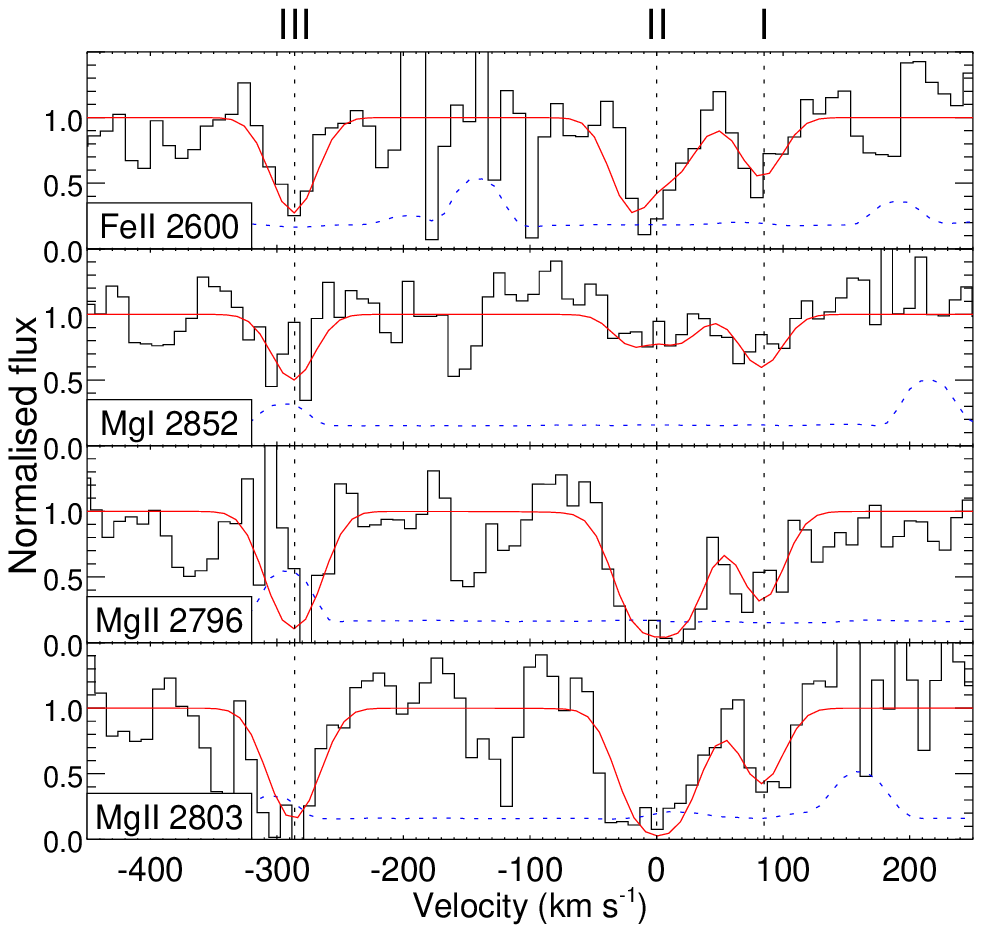}
      \caption{Fit of \ion{Mg}{I}, \ion{Mg}{II} and \ion{Fe}{II} lines of the intervening system at $z$=1.8005, with three velocity components.
              }
         \label{Fig1.8}
   \end{figure}

At a redshift of 1.8005 we detect a further intervening system through several absorption features produced by low ionisation transitions (\ion{Fe}{II} and \ion{Mg}{II}) as well as neutral \ion{Mg}{I}.  This  system is formed by three components with relative velocities of --286 km s$^{-1}$, 0 km s$^{-1}$ and +85 km s$^{-1}$ (see Fig.~\ref{Fig1.8} and Table~\ref{table:int2}).

The absorber at $z=1.8$ can be classified as a strong absorber (\ion{Mg}{II}$\lambda$2796 EW$_{\mathbf{rest}} > 1$ \AA, see Table~\ref{table:ewlines}). \citet{pro06} pointed out an excess (by a factor of $\sim 4$) in the frequency of strong \ion{Mg}{II}$\lambda$2796 absorbers in the line of sight of GRBs as compared to quasars. Recently, \cite{ver09}, using a sample of GRB afterglow spectra that doubles the redshift path of previous studies, confirmed this excess but only by a factor of $\sim2$. The comparison is done using the SDSS QSO survey, with a redshift range $0.37<z<2.27$ and a 6-$\sigma$ detection limit of the \ion{Mg}{II}$\lambda$2796 line. Following these criteria the redshift path of the X-shooter spectrum of GRB\,090313 for strong systems is $\Delta z=1.13$, giving a number density of 0.88. This number density is in agreement with that of 0.70$\pm$ 0.15 found by \cite{ver09}, whereas the number density of the strong \ion{Mg}{II} systems found along QSO lines of sight is only $0.28\pm0.01$ (Nestor et al. 2005).

\section{Discussion and conclusions}

In this paper we present the first spectrum of a GRB obtained with X-shooter, covering for the first time a range from optical to NIR simultaneously. Although observed during an early commissioning phase and under unfavourable conditions, the results allow us to have an optimistic view of X-shooter's potential.

For the GRB, we determine a redshift of 3.3736, at which we detect absorption features with three velocity components. From the values in Table~\ref{table:lines} and from Fig.~\ref{FigGRB} it is inferred that component II has the highest column densities, and we used it to calculate the redshift. Component I is the least ionised (high ionisation lines are only weakly detected) while in component III we detect mainly highly ionised species. Although not conclusive, this could indicate that component III is the closest to the GRB, while component I is the furthest. The relative velocities between the 3 different components are due to relative motion of the gas within the host galaxy, in principle unrelated to the GRB itself. The large values of [Si/Fe] measured in the GRB host galaxy may be indicative of enhanced abundance of Si (produced by extensive star formation) or a higher depletion level of Fe (evidence of dust). In fact, \citet{kan09} measured an extinction corresponding to $A_V$=0.34$\pm$0.15 for this burst, which as they note, is atypically large for such a high redshift event. They also note that GRB\,090313 is among the most intrinsically luminous optical afterglows detected. However, if we look at the EW$_{\mathrm{rest}}$ measured for the absorption features in this spectrum (see Table~\ref{table:ewlines}), we find that they are consistent with the typical values found in the sample of \citet{fyn09}. We note that, to our knowledge, some features such as \ion{Mg}{II}, \ion{Mg}{I} and \ion{Ca}{II} had not been observed before in GRB spectra at such high redshift. 

The surveys of strong  \ion{Mg}{II} (EW$_{\mathrm{rest}}$ $>$1~\AA) intervening absorbers along GRB lines of sight gave the surprising result of an excess of these systems compared to QSO lines of sight \citep{pro06,ver09}. We have measured the EW$_{\mathrm{rest}}$ of the \ion{Mg}{II}$\lambda$2796 line of both intervening systems, finding a combined value (coadd of the three components) of  1.35$\pm$0.10 $\AA$ for the $z$=1.80 and 0.50$\pm$0.03 $\AA$ for the $z$=1.96. The number density of strong \ion{Mg}{II} systems for this spectrum is ${\rm d}n/{\rm d}z = 0.88$, in agreement with the excess of a factor of $\sim 2$ compared to QSO lines of sight found by \cite{ver09}. The same authors also considered the statistics of weak systems (0.3~\AA$<$EW$_{\mathrm{rest}}$ $<$1~\AA) and conclude that their number density is in agreement with what is found along QSO lines of sight, as confirmed by \citet{tej09}. The S/N of the GRB\,090313 X-shooter spectrum is too low to cover a significant redshift path for a rest frame equivalent width limit of 0.3\AA, on the other hand this limit will be normally reached for the guaranteed time programme for observations of GRB afterglows that will be performed with X-shooter. This program will collect about 100 spectra within 3 years, hugely increasing the redshift path of the GRB surveys and therefore the significance of the statistics, and bringing useful information to explain the unexpected excess of strong absorbers. Moreover, X-shooter will cover simultaneously a larger redshift path compared to the instruments presently used for QSO and GRB afterglow spectroscopic observations, up to $\Delta z\sim 5$ (the total redshift path for strong \ion{Mg}{II} systems for the spectrum presented here is $\Delta z=2.1$). Therefore X-shooter QSO and GRB surveys will be able to investigate systematically the presence and nature of \ion{Mg}{II} systems up to a much higher redshift than currently done.

No significant emission lines have been detected at the redshift of the GRB or at any of the intervening systems. \citet{ber09} pointed out the presence of two galaxies near the afterglow (G1 and G2, see Fig.~\ref{FigAcq}). The brightest one (G1, $r$=15.6) has a known redshift of $z$=0.0235 and is located at 17\farcs8 from the afterglow, equivalent to 8.3 kpc projected at the redshift of the galaxy. We do not see any features at this redshift and, in particular, nothing is detected at the wavelength of the most prominent absorption features that can be seen in the SDSS spectrum of G1 \citep{aba09}, which are the \ion{Ca}{II}$\lambda$8500, 8544, 8664 and \ion{Na}{I}$\lambda$5891, 5897.  \ion{Ca}{II} features are not detected with 3-$\sigma$ limits of EW$_{\mathrm{rest}}<$0.6 $\AA$ and  \ion{Na}{I} with limits of EW$_{\mathrm{rest}}<$1.2 $\AA$. The closest galaxy, G2 at 2\farcs3, with $r$=21.6 does not have an identified redshift, so its relation with any of the observed absorption features will need further investigation.

To conclude, we point out some facts that show the potential that X-shooter has in the GRB field. Given a 10-$\sigma$ limiting magnitude (per spectral bin) of \textit{R}$\sim$21.0 with 1 hr exposure, it will be able to study 33\% (72\%) of long bursts 12 (2) hours after the onset \citep[using optical fluxes and decay indices from][]{nys09}. As an example, X-shooter would have been able to obtain a spectrum similar to or better than the one presented here of the $z\sim8.2$ GRB\,090423 \citep{tan09,sal09} during the first 24 hours and to follow the nearby GRB\,030329 for over a month. If we consider short GRBs, we expect to be able to study 3\% (17\%) within 12 (2) hours of the burst onset. Together with the rapid response mode available at VLT, that allows reaction times of just a few minutes, the possibilities increase. With its intermediate resolution, we will distinguish components with differential velocities of $\sim$30 km s$^{-1}$. As shown here, this resolution will allow to derive, through line fitting, abundances of the different element species, although with some limitations as compared to higher resolution spectrographs when fitting blended components or marginally saturated lines. Thanks to the wide wavelength coverage it will be capable of observing afterglows up to redshifts of $\sim$18, assuming that they exist (at redshift 18, Ly-$\alpha$ would lie at $\sim$23\,000~$\AA$). It will allow systematic studies of lines over wide redshift ranges (\ion{Mg}{II}$\lambda$2796, 2803 will be detectable from $z$=0.1 up to $z$=7.5) as well as measurements of ratios with widely separated lines. Furthermore, through the wide spectral coverage we will have a larger amount of lines to determine metallicities with better accuracy. It will also allow the study of dust extinction profiles in GRB environments. Thanks to the increased sensitivity compared to other instruments of similar resolution, we will be able to study feature variability with higher time resolution and up to later times. All this places X-shooter in a position to lead breakthrough advances in GRB research in the next years.

\begin{acknowledgements}
We dedicate this paper to the memory of Roberto Pallavicini, who greatly contributed to the success of X-shooter, but prematurely passed away before its first light.
AdUP acknowledges support from an ESO fellowship. This work is supported by ASI grant SWIFT I/011/07/0, by the Ministry of University and Research of Italy (PRIN MIUR 2007TNYZXL). The Dark Cosmology Centre is funded by the Danish National Research Foundation. We thank the anonymous referee for constructive comments.
\end{acknowledgements}

\bibliographystyle{aa}
\bibliography{090313_v9}

\end{document}